\documentstyle[twoside,fleqn,espcrc2,epsf,psfig]{article}
\newcommand{\AmS}{{\protect\the\textfont2
  A\kern-.1667em\lower.5ex\hbox{M}\kern-.125emS}}
  
\title{Exotica and the Confining Flux}

\author{J.~Kuti\address{Department of Physics, 
 University of California at San Diego,
 La Jolla, California 92093-0319}}

\begin{document}

\begin{abstract}
Recent developments in Quantum Chromodynamics (QCD) are reviewed
on three major topics where nonperturbative gluon excitations of the 
QCD vacuum and the physical properties of the confining flux play
a central role: (1) New lattice
results on the spectrum and wave functions of heavy 
${\rm Q}\overline{{\rm Q}}{\rm gluon}$ molecules, known as heavy
hybrids, will be discussed. 
(2) Recent advances on the
glueball spectrum in lattice QCD will be 
presented with some theoretical observations. 
(3) Progress in our understanding
of the nonperturbative internal structure of the confining flux and 
its excitation spectrum will be reported.
\end{abstract}

\maketitle

\section*{INTRODUCTION AND OUTLINE}
I will review here three topics with
considerable progress in phenomenologically
relevant and theoretically challenging areas of lattice QCD.
They are all related to nonperturbative gluon excitations of the 
QCD vacuum and the physical properties of the confining flux.
The limited scope of the talk does not
include discussions on some exotic objects, like
light hybrids~\cite{Kenway,Lacock} and dibaryons~\cite{Negele1}, or
glueball phenomenology~\cite{Kenway,Weingarten}, and some
other topics which were
reviewed, or presented elsewhere at the conference. 
The outline of the talk is as follows:
\newline
\indent\indent{\em 1. Heavy Hybrid Spectrum}
\newline
\indent\indent\indent\indent - Methods and Phenomenology
\newline
\indent\indent\indent\indent - Born-Oppenheimer Approximation
\newline
\indent\indent\indent\indent - NRQCD Hybrid Spectrum
\newline
\indent\indent\indent\indent - Results from the MILC Collaboration
\newline
\indent\indent{\em 2. Glueball Spectrum}
\newline
\indent\indent\indent\indent - Review of Recent Lattice Results   
\newline
\indent\indent\indent\indent - Composite Glueball Operators   
\newline
\indent\indent\indent\indent - Consituent Gluon Model   
\newline
\indent\indent{\em 3. The Confining Flux}
\newline
\indent\indent\indent\indent - Microscopic Vacuum Model   
\newline
\indent\indent\indent\indent - Casimir Scaling and Multiple Sources   
\newline
\indent\indent\indent\indent - Where Is the String Limit?   
\newline
\indent\indent\indent\indent - Flux Fission (String Breaking)
\vskip 0.1in
The chromoelectric flux with
static sources is the organizing theme of the selected topics.
Its ground state energy
explains the origin of the confining potential of heavy
${\rm Q}\overline{{\rm Q}}$ pairs. The excitation spectrum
of the confining (and confined) flux defines
hybrid ${\rm Q}\overline{{\rm Q}}$ potentials which bind quark-antiquark
pairs into
heavy hybrid ${\rm Q}\overline{{\rm Q}}{\rm gluon}$ molecules.
Higher excited flux states are unstable against
glueball emission requiring a detailed study
of the glueball spectrum.
The confined flux, when stretched, is expected to reach the limiting 
behavior of an effective QCD string in geometric string
variables which remains a topic of considerable theoretical interest.
The stretched flux can
also fission by quark-antiquark pair creation in the QCD vacuum with
important phenomenological implications.
Flux states with sources in higher color representations, or
with multiple color sources, like three-quark and
four-quark configurations, provide further information on 
flux formation in the confining vacuum.

\section{HEAVY HYBRID SPECTRUM}
I will review recent results on the
heavy hybrid spectrum from three complementary approaches, followed 
by a brief phenomenological note.
\subsection*{Methods and Phenomenology}
The first method is based on the Born-Oppenheimer expansion.
In this approach, the heavy hybrid meson is treated analogous to a diatomic
molecule:  the slow heavy quarks correspond to the nuclei and the fast
gluon field corresponds to the electrons~\cite{Hasenfratz,Mandula}.  
The first
step in the Born-Oppenheimer treatment is to determine the energy levels
of the glue (and light quark-antiquark pairs) as a function of the
heavy quark-antiquark separation, treating the heavy quark and antiquark
simply as spatially-fixed color sources.  Each such energy level defines
an adiabatic potential.  The quark motion is then restored by solving
the nonrelativistic Schr\"odinger equation using these potentials.
Conventional quarkonia arise from the lowest-lying potential; hybrid
quarkonium states emerge from the excited potentials. 
The leading Schr\"odinger kinetic energy term
and the omission of retarded transverse
gluons (in Coulomb gauge), radiated and reabsorbed by
the slowly moving quarks, could be corrected in a systematic expansion.

The starting point of the second method is the lattice formulation 
of the euclidean NRQCD framework with the heavy quark propagating 
according to a spin-independent nonrelativistic action~\cite{Lepage}. 
Retarded gluon effects are included, spin effects and corrections
to the leading kinetic term can be also added without difficulty.
I will report new NRQCD results which
show that gluon retardation effects are very small in the 
bottomonium system and its hybrid excitations confirming the success
of the leading Born-Oppenheimer approximation.

The third approach which was explored by the MILC collaboration~\cite{MILC2}
applies Wilson and clover lattice fermions in the mass range 
of the charmed quark 
to explore the hybrid excitations of the charmonium system. The 
heavy c quark is treated without nonrelativistic approximation and
all spin and  retardation effects are included.

\begin{figure}[htb]
\begin{center}
\epsfxsize=3.0in\epsfbox[96 72 596 720]{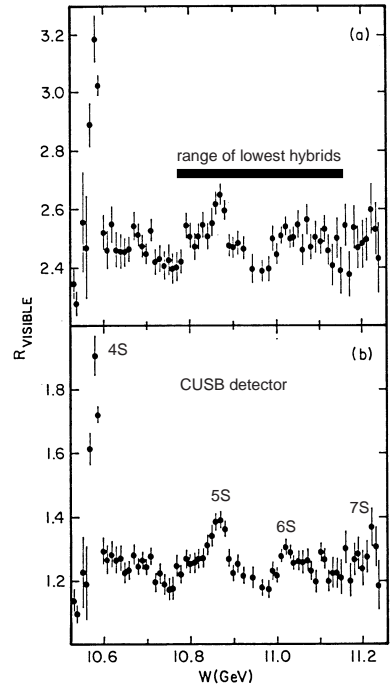}
\end{center}
\vskip -0.7in
\caption{The CUSB scan of the energy region above the ${\rm b}\bar{\rm b}$
threshold is shown~\cite{CUSB} 
together with the expected mass range of the lowest
heavy hybrid ${\rm b}\bar{\rm b}{\rm gluon}$ states. The conventional
interpretation of the resonance peaks corresponds to the radial
excitations of the $\Upsilon$ particle~\cite{Eichten}.}
\label{fig:fig02h}
\end{figure}

I should note that the phenomenology of heavy 
hybrids is very interesting.
Early results from the CUSB and 
CLEO collaborations~\cite{CUSB,CLEO} revealed a complex
resonance structure between the ${\rm b}\bar{\rm b}$ threshold and
11.2 GeV in  ${\rm e^+ e^-}$ annihilation
as shown in Fig.\ref{fig:fig02h}. This is precisely the energy
range where the lowest hybrid excitations are expected. Today
it would only take a very short run for the upgraded CLEO detector to rescan
this region with much higher resolution to find perhaps, for the first time,
gluon excitations in the hadron spectrum under experimentally and
theoretically rather well-controlled circumstances. This is an exciting
area where lattice work will remain the dominant method.
\vskip -0.2in
\subsection*{Born-Oppenheimer Approximation}
\begin{figure}[htb]
\begin{center}
\epsfxsize=2.8in\epsfbox[48 115 559 648]{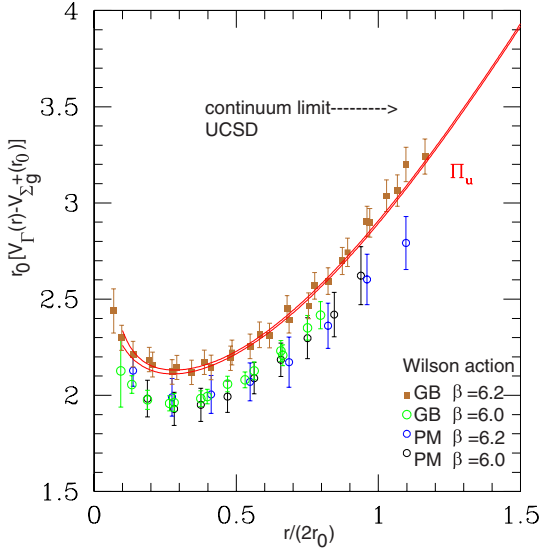}
\end{center}
\vskip -0.7in
\caption{Results from
mean-field improved anisotropic lattice actions~\cite{Juge,Juge1} 
are compared 
to recent~\cite{Bali1}(GB) and earlier~\cite{Michael}(PM) 
simulations with Wilson actions. The excited hybrid potential
has quantum numbers $\Gamma = \Pi_{\rm u}$. 
Consistency across  
Refs.~\cite{Juge,Juge1,Bali1} is excellent 
for small Wilson gauge coupling.}
\label{fig:fig03h}
\end{figure}

The lowest hybrid
potential which is depicted in Fig.\ref{fig:fig03h} corresponds 
to a gluon excitation with one unit of angular momentum projected along the
quark-antiquark (molecular) axis with ${\rm CP=-1}$ quantum numbers
(excited $\Pi_{\rm u}$ state above the $\Sigma_{\rm g}^+$ ground state
in spectroscopic notation). 
I will discuss the full spectrum of gluon excitations in Section 3.
Following Ref.~\cite{Sommer}, 
the physical scale ${\rm r_0}$  in Fig.\ref{fig:fig03h} is 
defined by the relation
$
{\rm [r^2 dV_{\Sigma^+_{\rm g}}(\vec{r})/dr]}_{\rm r=r_0}=1.65~.
$

\begin{figure}[tb]
\begin{center}
\epsfxsize=2.5in\epsfbox[60 185 534 593]{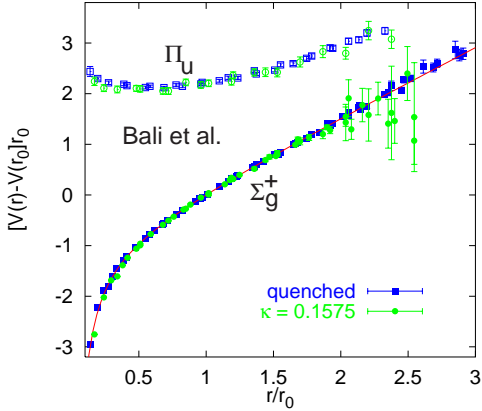}
\end{center}
\vskip -0.7in
\caption{Wilson quark loop effects are shown for the light quark 
mass range where 
${\rm m_\pi \sim m_\rho/2}$~\cite{Bali2}.}
\label{fig:fig04h}
\end{figure}

An important recent study~\cite{Bali2} shows in Fig.\ref{fig:fig04h} that
dynamical quark loop effects are not visible at larger distances,
out to 1.2 fm separation of the quark-antiquark pair in the
$\Sigma_{\rm g}^+$ and $\Pi_{\rm u}$ potentials. 
This further supports the accuracy and 
phenomenological relevance of the approach.

\begin{figure}[htb]
\vskip -0.2in
\begin{center}
\epsfxsize=2.3in\epsfbox[18 144 592 718]{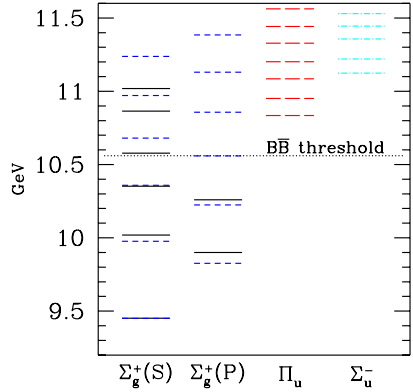}
\end{center}
\vskip -0.7in
\caption{The Schr\"odinger spectrum of the adiabatic hybrid potentials
is shown in comparison with the conventional S and P
states of the ${\rm b}\bar{\rm b}$ 
spectrum.}
\label{fig:fig05h}
\end{figure}

\begin{figure}[htb]
\begin{center}
\epsfxsize=2.5in\epsfbox[100 100 512 692]{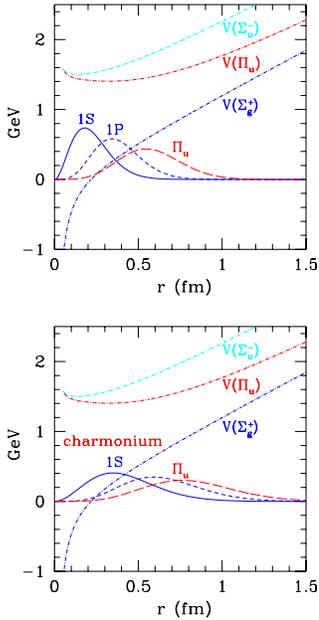}
\end{center}
\vskip -0.7in
\caption{The radial probability distribution
in the lowest $\Pi_{\rm u}$ hybrid state is shown in comparison
with conventional ${\rm b}\bar{\rm b}$ S and P states of the
bottomonium (upper box) and charmonium system.}
\label{fig:fig06h}
\end{figure}

The spectra of heavy hybrid states built on the two lowest
hybrid potentials ${\rm V_{\Pi_{\rm u}}(\vec{r})}$ and
${\rm V_{\Sigma^-_{\rm u}}(\vec{r})}$
are shown 
in Fig.\ref{fig:fig05h}. The most prominent feature of the
spectrum is the dense set of radial excitations in the
rather shallow hybrid potentials with level separations 
$\sim {\rm 200-300~MeV}$. 
The hybrid wavefunctions all
vanish at the origin due to the repulsive Coulomb core in the potential
and the special centrifugal term in the adiabatic Schr\"odinger equation
~\cite{Hasenfratz,Mandula}. The hybrid Schr\"odinger states are also
much more extended than the conventional quark-antiquark states, 
as depicted in Fig.\ref{fig:fig06h}.

Once the hybrid potentials are determined, the Born-Oppenheimer approach
yields the entire leading-order spectrum, in contrast to
direct NRQCD simulations.  However, the validity of the Born-Oppenheimer
approximation relies on the smallness of retardation effects. In addition,
spin-dependent corrections have to be included for precise
predictions.
Quantitative estimates on retardation effects 
comes from direct comparison of mass splittings
in the leading Born-Oppenheimer approximation with
those determined from NRQCD simulations.
\subsection*{NRQCD Hybrid Spectrum}
\begin{figure}[t]
\begin{center}
\epsfxsize=2.3in\epsfbox[44 116 549 648]{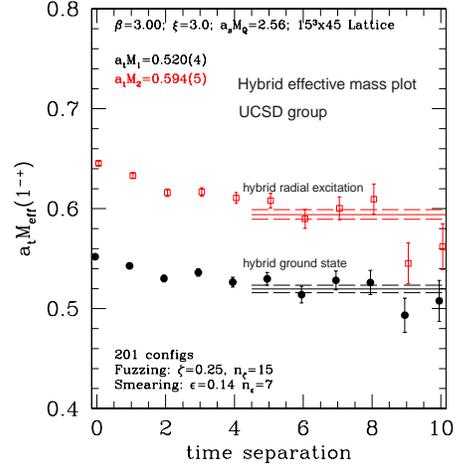}
\end{center}
\vskip -0.7in
\caption{Mass estimates are shown for the lowest hybrid $\Pi_{\rm u}$ state 
and its first radial excitation~\cite{Juge1}.}
\label{fig:fig07h}
\end{figure}

Recent results on heavy hybrid states were reported
using anisotropic coarse
lattices with mean-field improved action~\cite{Juge,Juge1},
and Wilson gauge action~\cite{Horgan,Glasgow}. 
In Ref.~\cite{Juge1}
the bare quark mass was taken to be ${\rm a_sM_b=2.56}$.
The so-called kinetic mass of the $\Upsilon$ was then determined from
its low-momentum dispersion relation.  Half of this mass was used
for the quark mass in the leading Born-Oppenheimer calculation. This
ensured that the $\Upsilon$ kinetic masses were identical in both
calculations. 
A typical effective mass plot for hybrid states
from the NRCQD run with improved action on asymmetric lattices
is shown in Fig.\ref{fig:fig07h}~\cite{Juge1}. 
The hybrid mass with Wilson gauge action~\cite{Horgan}
was found noticeably higher, but part of the disagreement
might be attributable to the different choice of setting the 
physical scale.
Results from Ref.~\cite{Juge1} are consistent
with earlier work~\cite{Juge,Glasgow} adding improved control on systematics
and statistics. 
One of the new features here is the clear signal
for the first radial hybrid excitation.
The statistical error on the mass of the
hybrid ${\rm b\bar b g}$ state is small
but the systematic study of lattice artifacts remains incomplete.
Assuming small lattice artifacts in the
simulation results, it was shown that
gluon retardation affects
the spin-averaged mass splittings by less than ten percent, 
validating the leading
Born-Oppenheimer approximation~\cite{Juge1}.

\subsection*{Results from the MILC Collaboration}
\begin{figure}[htb]
\begin{center}
\epsfxsize=3.0in\epsfbox[15 152 592 728]{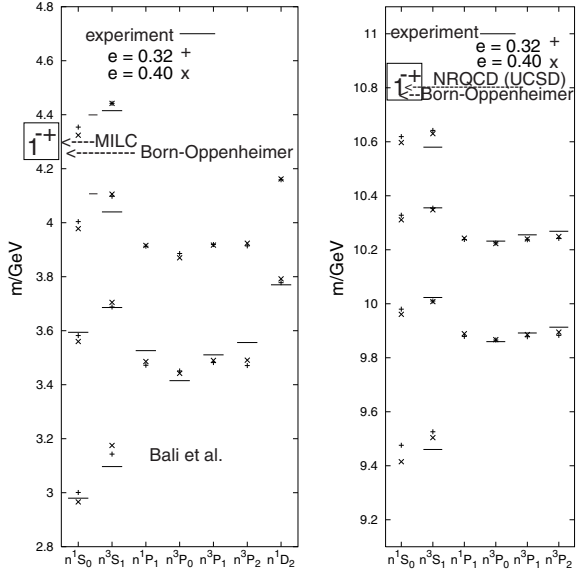}
\end{center}
\vskip -0.6in
\caption{The lowest hybrid states are shown in the
charmonium and bottomonium spectrum from
three different methods described in the text. The 
spin-dependent potential model spectrum 
is superimposed for comparison from Ref.~\cite{Bali3} where the notation
is explained.}
\label{fig:fig09h}
\end{figure}

The MILC collaboration repeated its original Wilson simulations 
of the hybrid meson spectrum~\cite{MILC1}
using the clover action~\cite{MILC2}. 
The new results, shown 
in  Fig.\ref{fig:fig09h}
for the $1^{-+}$ heavy hybrid mass,
are consistent with earlier MILC Wilson runs providing
an important check on charmed quark lattice artifacts.
Preliminary 
results for the wave function of the $1^{-+}$ state in Coulomb gauge were
also reported at the conference~\cite{MILC2}.
\section{GLUEBALL SPECTRUM}
In the perspective of ten years it is quite remarkable
how much is known now about the quenched glueball spectrum. Hopefully, it 
will not take another ten years before quark-antiquark
vacuum polarization effects are included systematically, with control
comparable to the quenched results.

\subsection*{\bf Review of Recent Lattice Results}

\begin{figure}[htb]
\begin{center}
\epsfxsize=2.5in\epsfbox[41 52 529 717]{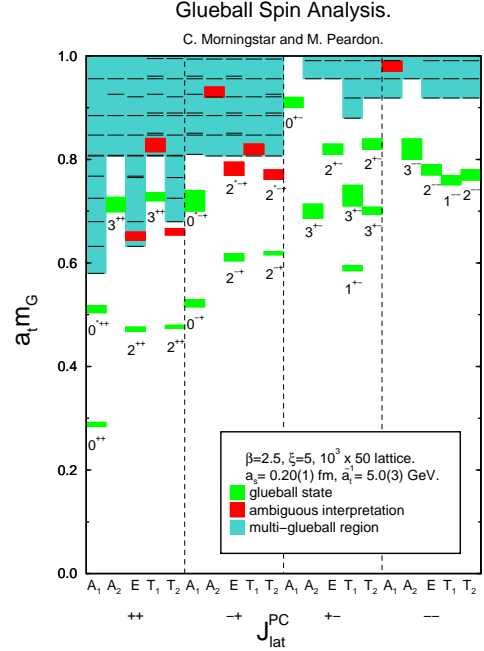}
\end{center}
\vskip -0.6in
\caption{The SU(3) lattice spectrum of glueball states in the quenched
approximation.}
\label{fig:fig01g}
\end{figure}

\begin{figure}[htb]
\begin{center}
\epsfxsize=2.5in\epsfbox[66 53 553 624]{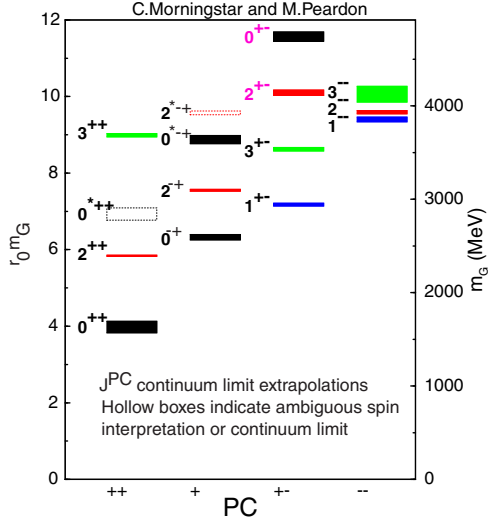}
\end{center}
\vskip -0.7in
\caption{Final spin analysis of the continuum glueball spectrum.}
\label{fig:fig02g}
\end{figure}

\begin{figure}[htb]
\begin{center}
\epsfxsize=2.5in\epsfbox[50 90 530 660]{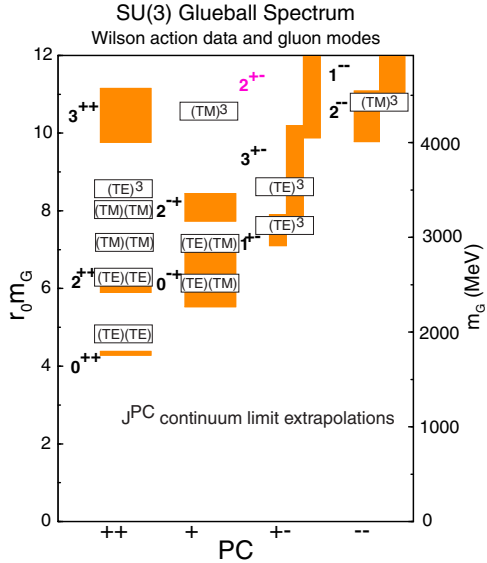}
\end{center}
\vskip -0.6in
\caption{The range of energy levels from several Wilson action 
runs is indicated by shaded areas. 
The constituent gluon spectrum (discussed in the text) is also displayed
with the occupied gluon modes.}
\label{fig:fig03g}
\end{figure}

The construction of improved
anisotropic lattice actions using mean-field improved
perturbation theory is discussed in Ref.~\cite{Morn1} with further pointers
to the literature. All recent results
reviewed here are based on this action which was designed 
to be $O(a_s^2)$ improved and to
include terms that directly couple links only on adjacent time slices 
to avoid spurious modes in the gluon propagator.
The hadronic scale parameter ${\rm r_0}$, defined earlier, is used
in the conversion of the measured glueball masses into physical units.
The method has been applied to the SU(3) glueball spectrum by
Morningstar and Peardon~\cite{Morn1} whose comprehensive lattice spin
analysis~\cite{Peardon,Morn2} is shown in Fig.\ref{fig:fig01g}.
Recently 
a detailed
analysis of the SU(2) glueball spectrum using very similar
methods was also presented~\cite{Trot1}.
The assignment of continuum spin
labels to lattice states in Fig.\ref{fig:fig01g} is based on transformation
properties of the glueball states
under the cubic point group, combined with parity and charge
conjugation operations~\cite{Morn1,Morn2}.
The continuum spin quantum numbers are not always fully resolved 
across cubic group representations (designated as ambiguous states).
The bottom of the shaded
area in each channel of Fig.\ref{fig:fig01g} indicates where the continuum
threshold of the lightest
two-glueball state is located.
Glueball states above the continuum threshold are not included
in the final spin analysis of the continuum spectrum
in Fig.\ref{fig:fig02g}.
The two lightest glueballs, the scalar and tensor,
have candidate resonances~\cite{Anis94,Balt85,Goda97}:
the $f_0(1500)$ and $f_J(1710)$ scalar states
and the $\xi(2230)$ tensor candidate. 
Both $f_0(1500)$ and 
$f_J(1710)$ are consistent with lattice mass predictions in the 
pure-gauge theory ~\cite{UKQCD93,Sext95,Morn1} of 1600 MeV with 
systematic errors of approximately 100 MeV.
The glueball spectrum from the Wilson action is
shown in Fig.\ref{fig:fig03g}, together with the spectrum from
a constituent gluon model. 

The SESAM collaboration~\cite{Bali97} has investigated the scalar and tensor 
glueballs on their large ensemble of unquenched (Wilson gauge 
and quark action) configurations. 
Only little change has been found in the masses of these states 
as the sea quark mass runs down
from the strange quark mass to 
($m_\pi/m_\rho \approx 0.5$) but enhanced finite-volume
dependence was noticed which is perhaps attributable
to dynamical quark effects. 

\subsection*{Composite Glueball Operators}

In a confining theory it is difficult to relate physical glueball states
to fundamental gluon fields. There is, however, a qualitative expectation
that higher dimensional interpolating operators will create
higher mass states~\cite{Minkowski,Bjorken,Jaffe}. This rule works
well for light mesons and baryons and now I will apply it to 
the glueball spectrum.

Local composite operators of the gluon field
act as interpolating fields for glueball
states. For example, the state with $0^{-+}$ quantum numbers
is created from the physical vacuum by the operator
$\vec{{\rm E}}_{{\rm a}}\cdot\vec{{\rm B}}_{{\rm a}}$ 
(a=1,2,...,8 stands for color summation)
so that
$\langle 0^{-+}|\vec{{\rm E}}_{{\rm a}}\cdot
\vec{{\rm B}}_{{\rm a}}| {\rm vacuum}\rangle \neq 0$. 
It is not difficult to determine the quantum numbers
of all the glueball states which are generated
by the lowest dimensional (dim=4,5,6) operators. 
A small representative sample is listed in 
Table~\ref{table:Operator}.
\vskip -0.1in
\begin{table}[htb]
\setlength{\tabcolsep}{1mm}
\caption{Sample of dim=4,6 operators which create various
${\rm J}^{{\rm PC}}$ quantum numbers.
\label{table:Operator}}
\begin{center}
\begin{tabular}{c|l|c} \hline
 dimension & ${\rm J}^{\rm PC}$ & Operator\\ \hline\hline
 
 4  & $0^{-+}$ & $\vec{{\rm E}}_{{\rm a}}\cdot\vec{{\rm B}}_{{\rm a}}$   \\
 4  & $0^{++}$ & $\vec{{\rm E}}_{{\rm a}}^2 - \vec{{\rm B}}_{{\rm a}}^2$ \\
 4  & $2^{++}$ & $\vec{{\rm E}}^i_{{\rm a}}\vec{{\rm E}}^j_{{\rm a}} +
                  \vec{{\rm B}}^i_{{\rm a}}\vec{{\rm B}}^j_{{\rm a}} -
                  \frac{1}{3}\delta^{\rm ij}(\vec{{\rm E}}_{{\rm a}}^2 +
                  \vec{{\rm B}}_{{\rm a}}^2)$ \\  
\hline 
 6 &  $0^{-+}$ & ${\rm f}_{\rm abc}(\vec{{\rm E}}_{\rm a}
          \times\vec{{\rm E}}_{\rm b})\cdot \vec{{\rm E}}_{\rm c} $\\ 
 6 &  $1^{-+}$ &  ${\rm f}_{\rm abc}(\vec{{\rm B}}_{\rm a}
                  \cdot\vec{{\rm E}}_{\rm b}) \vec{{\rm B}}_{\rm c} $\\  
 6 &  $2^{-+}$ &  ${\rm f}_{\rm abc}(\vec{{\rm E}}_{\rm a}
          \times\vec{{\rm E}}_{\rm b})^i \vec{{\rm E}}_{\rm c}^j + ... $\\
\hline
\end{tabular}
\end{center}
\vskip -0.2in
\end{table}
In classifying the lowest glueball excitations according their quantum
numbers, Morningstar and Peardon report 15+5 states
where the last five
are labelled as questionable. The lowest dimensional
operators with dim=4,5,6 create states whose quantum numbers
match onto the observed spectrum nearly perfectly. This is quite remarkable,
given the simplicity of the argument.

\subsection*{\bf Constituent Gluon Model}
The notion of consituent
gluons stems from the general idea of quasi-particles in some effective QCD
mean field theory of hadrons. In such a description the 
fundamental gluon field is replaced by Hartree modes of 
some constituent gluon field with residual perturbative
interactions inside glueballs. 
It is hoped, for example, that the 1/N expansion
might eventually lead to a similar Hartree picture.
I will identify the Hartree states with free gluon modes
in a bubble (bag) with confining boundary conditions~\cite{Close,Hansson}.
Admittedly speculative, the model in its simplest form
leads to detailed predictions on the glueball spectrum.

\begin{figure}[htb]
\begin{center}
\epsfxsize=2.7in\epsfbox[56 -8 669 785]{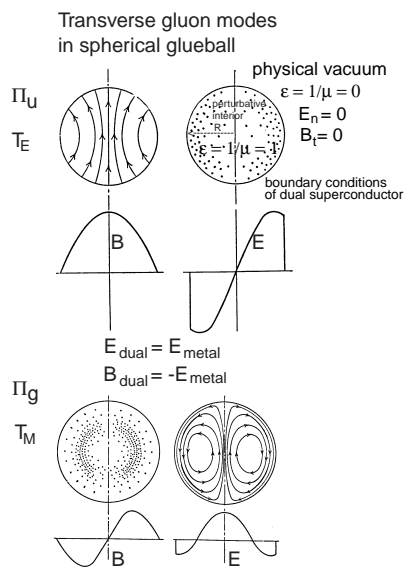}
\end{center}
\vskip -1.0in
\caption{Hartree states of TE and TM gluon modes.}
\label{fig:fig06g}
\end{figure}

We will picture the QCD vacuum as a diaelectric medium (dual superconductor)
with vanishing chromoelectric permeability and infinite magnetic
permeability. Glueballs are 
floating bubbles in the physical vacuum (Fig.\ref{fig:fig06g})
and their perturbative interior is occupied by constituent TE and TM gluon 
modes appropriate for the confining boundary conditions of a dual 
superconductor~\cite{Close,Hansson}. 
The energy density in the physical vacuum (which is the required vacuum
energy to create the bubble) is given by the well-known expression
$
{\rm B} = -\frac{9}{32\pi}\alpha_{{\rm s}}
\langle 0|{\rm F}^2_{\mu\nu}|0\rangle
$
from the trace anomaly of 
the energy-momentum tensor~\cite{Shifman}.
The quantum numbers and energies of the lowest gluon modes are given
in Table~\ref{table:gluons}
and the fields of the TE and TM modes are depicted 
in Fig.\ref{fig:fig06g}.
\begin{table}[htb]
\setlength{\tabcolsep}{1mm}
\caption{Lowest TE and TM gluon modes in spherical bubble with energies
${\rm x}_{{\rm i}}/{\rm R}$ for the ${\rm i}^{{\rm th}}$ mode.
\label{table:gluons}}
\begin{center}
\begin{tabular}{l|l|l|l} \hline
 l & mode & ${\rm J}^{\rm P}$ & ${\rm x}_{{\rm i}}$ \\ \hline\hline
 1 & TE   & $1^+$             & 2.744 \\
 2 & TE   & $2^-$             & 3.96  \\
 1 & TM   & $1^-$             & 4.49  \\ \hline 
\end{tabular}
\end{center}
\vskip -0.2in
\end{table}
The energy of the localized glueball is calculated from
\begin{equation}
{\rm E}_{{\rm glueball}} = 
\frac{4\pi}{3}{\rm R}^3 {\rm B} + \sum_{{\rm i}} {\rm n}_{{\rm i}}
{\rm x}_{{\rm i}}/{\rm R} - {\rm c} \frac{\alpha_{{\rm s}}}{{\rm R}}
\lambda_1^{\rm a}\lambda_2^{\rm a}\vec{{\rm S}}_1\vec{{\rm S}}_2,
\label{eq:Eglue}
\end{equation}
where ${\rm n}_{{\rm i}}$ is the occupation number of the
${\rm i}^{{\rm th}}$ gluon mode and the numerical constant c is close to 1/4.
The last term in Eq.~(\ref{eq:Eglue}) represents the interaction energy
of the gluon modes. I chose ${\rm B}^{1/4} = 280~{\rm MeV}$ for
the vacuum energy density which is appropriate
for heavy quark spectroscopy and the hybrid potentials~\cite{Kuti}
(${\rm B}^{1/4} \sim 250~{\rm MeV} - 350~{\rm MeV}$
is the preferred range from the evaluation of the 
gluon condensate~\cite{Shifman}). 
The mass of the glueball has to be corrected for the localization
of the center-of-mass,
\begin{equation}
{\rm M}^2_{{\rm glueball}}
= {\rm E}^2_{{\rm glueball}} - \sum_{{\rm i}} {\rm n}_{{\rm i}} 
{\rm x}^2_{{\rm i}}/{\rm R}^2,
\label{eq:Emass}
\end{equation}
where the second term is an estimate of the kinetic
energy $\langle \vec{{\rm P}}^2\rangle$. The energies and radii
of the various glueball states
are determined from the minimization of Eqs.(\ref{eq:Eglue},\ref{eq:Emass})
with respect to R. The spectrum is displayed in Fig.\ref{fig:fig03g}
with $\alpha_{\rm s}({\rm 1/R}) = 0.5$ where ${\rm R} \approx {\rm 0.5~fm}$.

The old bag calculations of the glueball
spectrum did not work. They led to very low glueball 
masses with wrong spin splittings for two reasons: 
the wrong value of B
and an unacceptably large effective $\alpha_{{\rm s}}$ were used
simultaneously
from light hadron spectroscopy~\cite{Close,Hansson}. 
I have nothing to say about
light hadrons which are intimately
connected with the chiral condensate of the vacuum. 
In contrast, the dual superconductor picture 
gives consistent phenomenology for so many aspects
of the microscopic vacuum structure in the gluon sector 
of the theory~\cite{Kuti}. 
It remains a mystery why this simple caricature of
glueball states works so well in describing the quenched
spectrum. Only a microscopic theory of the vacumm could shed light on this
question.

\section{THE CONFINING FLUX}

I will discuss first an interesting
microscopic vacuum model
to illustrate the generic features of the confining flux. The
Lagrangian of the same model was also used in two contributions 
to the conference
in the study of flux fission (string breaking)~\cite{Wittig,Sommer1}.

\subsection*{Microscopic Vacuum Model}

We will investigate the response  of the microscopic 
monopole vacuum to Wilson loops in
the 2+1 dimensional Georgi-Glashow model as
the simplest example of quark confinement~\cite{Polyakov}.
In the Higg vacuum of the theory the SU(2) gauge symmetry is broken
to U(1) with a neutral and massless gauge boson ${\rm A}_\mu$.
The massive Higgs boson is frozen out from the low energy
theory, together, in the weak coupling limit  
${\rm g}^2/{\rm m}_{\rm W} \ll 1$, with
the two massive gauge bosons ${\rm W}^{\pm}$ 
where the dimensional gauge coupling g sets the physical scale.
The euclidean vacuum is dominated by a dilute gas of instantons
which are 't Hooft-Polyakov monopoles
interacting with a magnetic Coulomb force.

The monopole vacuum responds to external U(1) charges 
by flux formation. This is calculable from the
response of the vacuum to an external magnetic potential $\eta$,
given by the partition function
\begin{eqnarray}
\lefteqn{ Z(\eta) = 
\int{\mathcal D}\chi {\rm exp}\left\{ -\frac{{\rm g^2}}{{\rm 16\pi^2}} 
\int \right.  } \nonumber \\ 
& & \left. \left[ \frac{1}{2}(\nabla (\chi - \frac{1}{2}\eta))^2 
      -{\rm M^2} {\rm cos}\chi \right]{\rm d^3r}  \right\}~,
\label{eqn:Partit2}        
\end{eqnarray}
in the sine-Gordon field  representation of 
the monopole plasma~\cite{Polyakov}.
The sine-Gordon mass M is expressed in terms of the fugacity $\zeta$
of the monopole plasma by
${\rm M^2} = {\rm 32\pi^2} \zeta/{\rm g^2}$ where 
$\zeta \sim {\rm exp}({\rm -const\cdot m_W/g^2})$
is exponentially small in the weak coupling limit so that we have a
dilute monopole system.
A pair of static sources corresponds to an external magnetic potential
$\eta_{\rm wl}$ which is generated by an 
imaginary current ${\rm I=i\cdot g/2}$ running around the Wilson loop
immersed in the monopole plasma,
$
  \langle {\rm exp}({ {\rm i\int A_\mu d x_\mu}}) \rangle = 
  {\rm Z(\eta_{\rm wl})}/{\rm Z(0)} ~.
$
In the weak coupling limit the semiclassical expansion is applicable
(Debye-H\"uckel theory), and flux formation in leading order
is described by the solution of the sine-Gordon equation 
in the presence of the current loop. 
An exact sine-Gordon
soliton solution at infinite separation of the sources,
\begin{eqnarray}
  \chi_{\rm soliton} = \left\{ 
         \begin{array}{rll}
            {\rm  4arctg}({\rm exp(-M y)}) &,  &{\rm y > 0}\\
            {\rm -4arctg}({\rm exp(+M y)})  &,  &{\rm y < 0}
          \end{array}
                       \right.
\label{eq:sine-Gordon}                       
\end{eqnarray}
describes a confining flux running along the x direction, 
with the transverse
profile given by $\chi_{\rm soliton}$ in the y-direction.  
This solution in 2+1 dimensional
Minkowski space-time corresponds to the electric flux profile 
\[
  \vec{\rm E}_{\rm Minkowski} = \left(
-\frac{{\rm gM}}{{\rm 2\pi}}\frac{1}{{\rm cosh(My)}}, 0
\right)
\]
where the width of the flux which contains more than 90 percent
of the flux energy is approximately 4M,
set by the Debye-H\"uckel (sine-Gordon) mass ${\rm M}$. The 
asymptotic value of the string tension is given by $\sigma = 
\frac{\rm g^2}{\rm 2\pi^2}{\rm M}$.
                   
\begin{figure}[htb]
\begin{center}
\epsfysize=3.0in\epsfbox[0 66 476 761]{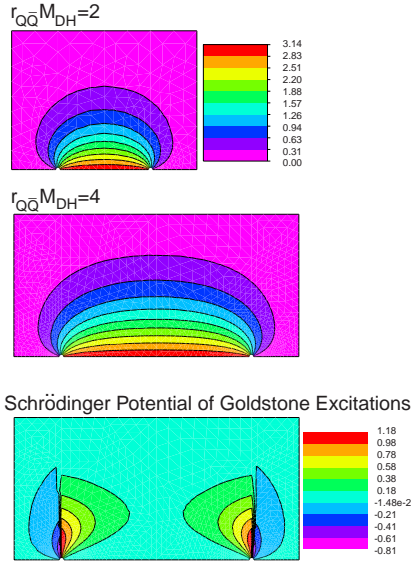}
\end{center}
\vskip -0.4in
\caption{Flux profiles at small and intermediate separation of
the sources. The bottom part is a contour plot of distortions
on the gutter shaped effective Schr\"odinger potential as discussed
in the text.}
\label{fig:fig01f}
\end{figure}

\begin{figure}[htb]
\begin{center}
\epsfysize=2.0in\epsfbox[48 109 528 750]{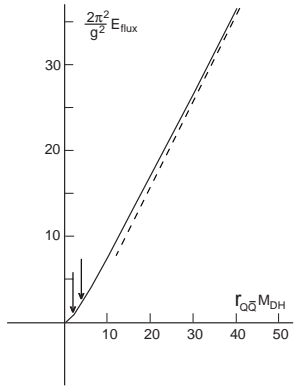}
\end{center}
\vskip -0.7in
\caption{Flux energy with arrows indicating the flux configurations
of Fig.\ref{fig:fig01f}. The slope of the dashed line is the exact
string tension of the infinitely long flux.}
\label{fig:fig02f}
\end{figure}

At finite separation of the sources we solve the 
sine-Gordon equation numerically~\cite{Juge2} with 
typical solutions depicted in Fig.\ref{fig:fig01f}.
We find a growing and stretching bubble in the 
vacuum crossing
from oblate to  prolate shape and  asymptotically converging to
the shape and intrinsic profile of the exact solution of
Eq.~(\ref{eq:sine-Gordon}) between the two fixed sources but not
close to them. The vortex energy is shown in 
Fig.\ref{fig:fig02f}.
When the separation of the sources becomes comparable
to the width of the flux, the rapid onset of
linear behavior in the flux energy, well before the asymptotic form
of the flux is reached, is a  
remarkable feature of Figs.\ref{fig:fig01f},\ref{fig:fig02f}. 
It is likely that a rather similar situation occurs in QCD
where we would like to answer the question: What sets the scale
of string formation?

\subsection*{Casimir Scaling and Multiple Sources}
Several interesting tests can be done in QCD on the flux formation mechanism
without looking into the details of the excitation spectrum.
One can excite the color sources at the two ends by
putting the quark and the antiquark into higher representations of
the SU(3) color group, or investigate flux formation
with several quark sources. 

Casimir scaling was a 
much discussed topic at the conference: according to folklore
the string tension should be proportional to the
value of the Casimir operator in the color
group representation. 
Deldar reported results for the
flux energy in the sextet and octet color
representations of the static quark-antiquark pair ~\cite{Deldar}.
A linearly rising potential with
Casimir scaling was observed in the approximate range of 0.5 fm
to 1.2 fm source separation, even for octet sources
where gluon screening is expected to split the flux asymptotically.
This is in agreement with earlier findings~\cite{Trot2,Campbell}.

Although it is popular to interpret Casimir scaling at intermediate
${\rm Q}\overline{\rm Q}$-separation as a test of microscopic
confinement mechanisms (Z(N) flux, monopoles, etc.), 
the test remains incomplete without a 
more detailed microscopic understanding of flux formation
and its dependence on color representations~\cite{Greensite}.

\begin{figure}[htb]
\begin{center}
\epsfxsize=2.0in\epsfbox[62 185 529 593]{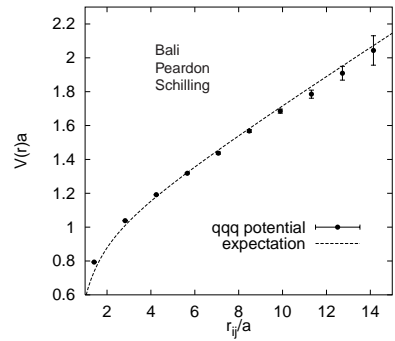}
\end{center}
\vskip -0.4in
\caption{3Q potential in Y-configuration.}
\label{fig:fig07f}
\end{figure}

\begin{figure}[htb]
\begin{center}
\epsfxsize=2.0in\epsfbox[108 177 505 615]{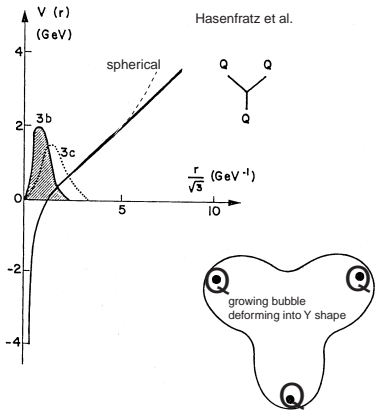}
\end{center}
\vskip -0.7in
\caption{3Q bag potential in Y-configuration. The dashed
line indicates spherical approximation.}
\label{fig:fig08f}
\end{figure}

In some interesting work in progress~\cite{Bali3}, the potential
energy of the 3Q system was studied. Anticipating the formation 
of a Y-shape string at large separation, the potential is plotted
in Fig.\ref{fig:fig07f} as a function of inter-quark
separation when the three quark sources are located in a plane
at equal angles.
The lesson from the growing and stretching bubble is that the rapid
onset of the linear potential cannot be interpreted
as evidence for the early formation of a Y-shape string. Earlier
work in the bag model illustrates that a very similar potential energy
shape develops precociously before string formation~\cite{Hasenfratz},
as depicted in Fig.\ref{fig:fig08f}.

Interesting new results were also reported at the conference on 4Q flux
configurations~\cite{Pennanen} which hopefully will advance our 
understanding of flux formation in the confining vacuum.

\subsection*{Where Is the String Limit?}

The confining flux is represented by the quantum sine-Gordon soliton
in the monopole vacuum. Excitations of the flux are given by
the spectrum of the fluctuation operator 
$
{\mathcal M} = -\nabla^2 + {\rm U^{"}(\chi_{soliton})}
$
where ${\rm U(\chi)}$ is the field potential energy of the sine-Gordon field.
The spectrum of the fluctuation operator ${\mathcal M}$ is determined
from a two-dimensional Schr\"odinger equation at finite separation of
the sources. 
In the limit of asymptotically large
${\rm r}_{\rm Q\overline {Q}}$, 
the equation becomes separable in the longitudinal x and transverse
y coordinates.
The transverse spectrum is in close analogy with the quantization of the
one-dimensional classical sine-Gordon soliton.
There is always a discrete zero mode in the spectrum which is 
enforced by translational invariance in the transverse direction. 
Scattering states of massive gauge bosons (``glueballs") on the
confining flux are separated by the mass gap M of the sine-Gordon
theory.

Plane wave excitations with momentum k along
the longitudinal direction, when combined with the zero mode
of the transverse spectrum, correspond to  Goldstone excitations
along the infinitely long flux. 
This spectrum is mapped into the spectrum of a string in the semiclassical
quantization procedure~\cite{Luescher1}. In string theory, the string
with fixed ends is described by Dirichlet boundary
conditions which would correspond to
closing down on the two ends
the long effective Schr\"odinger potential well with two infinite
potential walls (a gutter between infinite walls). 

At finite ${\rm r}_{\rm Q\overline {Q}}$, this picture, however
is grossly distorted. Far from the sources and from the main axis,
the effective Schr\"odinger potential well always has a finite height M. 
Therefore the gutter should be closed by two walls of finite
height M which are penetrable and distort the Goldstone spectrum.
In addition,
as depicted in Fig.\ref{fig:fig01f}, the potential well is further
deformed close to the sources. 
As a consequence, at smaller separation of the sources,
the fluctuation spectrum is not string-like at all.
At large separations, however,
the fluctuation spectrum of the quantum soliton (confining
flux) will smoothly deform with growing ${\rm r}_{\rm Q\overline {Q}}$
into the expected Goldstone spectrum. Based on the above 
picture this is not expected before the length of the flux exceeds several
times the width of the flux.

In QCD the robust features of the Goldstone spectrum of the confining
flux are described by an effective string theory where the
two-dimensional
vector $\vec{\xi}$ measures the deviations of the flux from
its straight line position~\cite{Luescher2}.
The effective low energy expansion is based on the assumption
that massive intrinsic excitations of the flux are decoupled
at low energies from the low-energy string-like excitations.
The effective
Lagrangian of this description has two important symmetries:
Euclidean invariance in the world sheet variables and O(2) translations
and rotations in the $\vec{\xi}$ field variable. The second 
symmetry requires
that ${\mathcal L}_{\rm eff}(\vec{\xi})$ can only depend on the
derivatives of $\vec{\xi}$,
$ {\mathcal L}_{\rm eff} = \frac{1}{2}\alpha\partial_\mu\vec{\xi}\cdot
\partial_\mu\vec{\xi} + ...$,
where the dots represent terms with four, or more derivatives
and $\alpha$ is a dimensional constant.
For low frequency excitations the first term in  
${\mathcal L}_{\rm eff}(\vec{\xi})$ dominates all the higher
dimensional operators. 

The Gaussian effective Lagrangian of the fluctuating flux describes
massless Goldstone excitations which are associated with the restoration
of Euclidean translations and rotations due to the formation
of the confining quantum flux. The full non-linear realization of euclidean
symmetries on $\vec{\xi}$ would amount to the summation of 
higher derivative operators requiring special coefficients and resulting
in the Nambu-Goto action,
\begin{equation}
 {\mathcal L}_{\rm eff} = \alpha \sqrt{ {\rm 1}+\partial_\mu\vec{\xi}\cdot
\partial_\mu\vec{\xi} } =  
 \frac{1}{2}\alpha\partial_\mu\vec{\xi}\cdot
\partial_\mu\vec{\xi} + ...
\end{equation}
This is in analogy with the nonlinear
chiral Lagrangian of massless pions. However, the special higher
derivative terms in the expansion of the Nambu-Goto action will compete
with other higher dimensional operators (rigidity,
torsion, and some other properties of the confining flux). It
is, therefore, difficult to believe that the Nambu-Goto action 
would play any useful role in the description of the excitation spectrum
of the confining flux.

The ground state energy of a string of length L with clamped ends
is given by 
\begin{equation}
{\rm E_0(L)} = 
\sigma\cdot {\rm L} - \frac{\pi}{{\rm 12L}} +...
\end{equation}
in 1-loop approximation where $\sigma$ designates
the string tension and the second term represents the Casimir
energy of zero-point fluctuations~\cite{Luescher2}. 
The excitation energies above the
ground state have the simple string spectrum, 
\begin{equation}
{\rm E_n = E_0} 
+ {\rm n}\cdot\frac{\pi}{\rm L}, ~~~~~{\rm n=1,2,3,...}
\end{equation}
It was repeatedly claimed by lattice practitioners 
that the Casimir
term was seen in the static quark-antiquark potential
(string ground state) for less than two fermi
separation, interpreted as strong evidence for string formation
in QCD. I will show now that this claim was intrinsically flawed.

In QCD we do not have the underlying microscopic theoretical picture
of the confining vacuum, but we expect to learn about it
by the determination
of the excitation spectrum of the confining flux in lattice simulations.
The first step in determining the spectrum of the stationary states
of the gluon field in the presence of the static quark-antiquark
pair, fixed in
space some distance ${\rm r}_{{\rm Q}\overline{\rm Q}}$ apart, is to 
classify the energy levels in terms of the
symmetries of the problem~\cite{Juge1}.  Such a system 
has cylindrical symmetry about
the axis $\hat{\bf r}$ passing through the static quark and the antiquark
(the molecular axis).  The total angular momentum of the gluons is not
a conserved quantity, but its projection onto the molecular axis
is and can be used to label the energy levels of the gluon field. 
We adopt the standard notation from the physics of diatomic molecules
and use $\Lambda$ to denote the magnitude of the eigenvalue of the projection
$\vec{J_g}\!\cdot\hat{\bf r}$ of the total angular momentum $\vec{J_g}$
of the gluon field onto the molecular axis $\hat{\bf r}$. The capital Greek
letters $\Sigma, \Pi, \Delta, \Phi, \dots$ are used to indicate states
with $\Lambda=0,1,2,3,\dots$, respectively.  
The energy of the gluon field is unaffected by reflections in a plane 
containing the molecular
axis; since such reflections interchange states of opposite handedness,
given by the sign of $\vec{J_g}\!\cdot\hat{\bf r}$, 
such states must necessarily be degenerate
($\Lambda$ doubling).  However, this doubling does not apply to the
$\Sigma$ states; $\Sigma$ states which are even (odd) under a reflection
in a plane containing the molecular axis are denoted by a
superscript $+$ $(-)$. 
The combined operations of
charge conjugation and spatial inversion about the midpoint between the
quark and the antiquark is also a symmetry and its eigenvalue is denoted by
$\eta_{CP}$.  States with $\eta_{CP}=1 (-1)$ are denoted
by the subscripts $g$ ($u$).  Hence, the low-lying
levels are labelled $\Sigma_g^+$, $\Sigma_g^-$, $\Sigma_u^+$, $\Sigma_u^-$,
$\Pi_g$, $\Pi_u$, $\Delta_g$, $\Delta_u$, and so on.  For convenience,
we use $\Gamma$ to denote these labels in general.

\begin{figure}[htb]
\begin{center}
\epsfysize=2.5in\epsfbox[25 52 545 573]{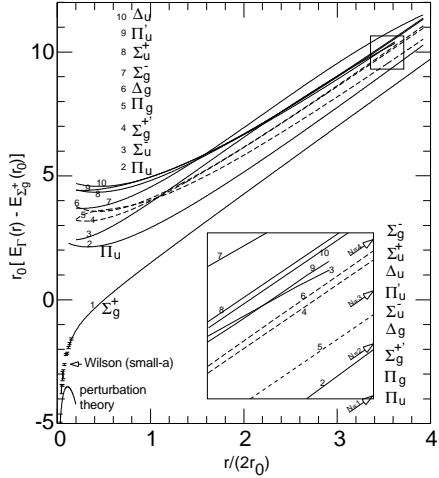}
\end{center}
\vskip -0.7in
\caption{The excitation spectrum of the confining flux. The energy
levels of N=1,2,3,4 Goldstone excitations are shown in the insert.}
\label{fig:fig03f}
\end{figure}

The continuum-limit extrapolations~\cite{Juge1} 
are shown in Fig.\ref{fig:fig03f}.
The ground state $\Sigma_g^+$ is the familiar static-quark potential.
A linearly-rising behaviour dominates the $\Sigma_g^+$ potential once
$r$ exceeds about 0.5 fm and we find no deviations from the linear form
up to 4 fm.  The lowest-lying excitation is the $\Pi_u$.  There is
definite evidence of a band structure at large
separation: the $\Sigma_g^{+\prime}$,
$\Pi_g$, and $\Delta_g$ form the first band above the $\Pi_u$;
the $\Sigma_u^+$, $\Sigma_u^-$, $\Pi_u^\prime/\Phi_u$, and $\Delta_u$
form another band.  The $\Sigma_g^-$ is the highest level at large 
${\rm r}_{\rm Q\overline {Q}}$.
This band structure breaks down as the separation
of the sources decreases below 2 fm. In particular,
two levels, the $\Sigma_g^-$ and $\Sigma_u^-$, drop far below their
degenerate partners as the separation between sources becomes small.  
Note that for 
${\rm r}_{\rm Q\overline {Q}}$ above 0.5 fm,
all of the excitations shown are stable with respect to glueball decay.
As the separation of the sources decreases below 0.5 fm, 
the excited levels eventually become
unstable as their gaps above the ground state $\Sigma_g^+$ exceed the
mass of the lightest glueball.

The level orderings and approximate degeneracies
of the gluon energies at large separation match those expected
of the Goldstone modes.  However, the precise Goldstone gap behaviour is not
observed.  The two
$\Sigma^-$ states are in violent disagreement with expectations from
a fluctuating string.  Note also that the results clearly disagree
with the energy spectrum of a Nambu-Goto string naively 
applied in four continuous space-time dimensions.  

These results are rather surprising and cast serious doubts
on the validity of treating glue in terms of a fluctuating string
for quark-antiquark separations less than 2 fm.  Note that such a
conclusion does not contradict the fact that the $\Sigma_g^+$ 
ground state energy
rises linearly for ${\rm r}_{\rm Q\overline {Q}}$ 
as small as 0.5 fm.  A linearly-rising term is not
necessarily indicative of a string as we have seen in earlier
examples.
For ${\rm r}_{\rm Q\overline {Q}}$
greater than 2 fm, there are some tantalizing signatures of
Goldstone mode formation, yet significant disagreements still remain.
To what degree these discrepancies can be explained in terms of
distortions of the Goldstone mode spectrum arising from the spatial
fixation of the quark and antiquark sources (clamping effect) 
is currently under investigation~\cite{Juge2}. 

For reasons explained here I remain puzzled and reserved
on reported results that
large Wilson loops in lower dimensional models agree 
with the predictions of a fluctuating string~\cite{Gliozzi} even for
relatively small separation of the sources. 

\subsection*{Flux Fission (String Breaking)}
The confining flux is expected to 
fission into a pair of static mesons, also known as string breaking, 
when its energy is large enough to
pair produce light dynamical quarks.
At large separation of the color sources the static potential, as
calculated from Wilson loops, 
describes the force between static 
mesons which are color sources screened by light quark
fields (like the static B meson in the infinite b-quark mass
limit).
We expect the linearly rising potential to asymptotically cross
over into Yukawa form controlled by the 
lowest mass exchange (pion). 
The attractive, or repulsive nature
of the Yukawa force will depend on the spin-isospin quantum
numbers of the static ${\rm B}\overline{\rm B}$ pair which
is not taken into account in Wilson loop operators. 
The crossover
range is expected to occur at a separation 
where the energy of the confining
flux exceeds twice the static meson energy.

\begin{figure}[htb]
\begin{center}
\epsfxsize=2.5in\epsfbox[75 307 572 665]{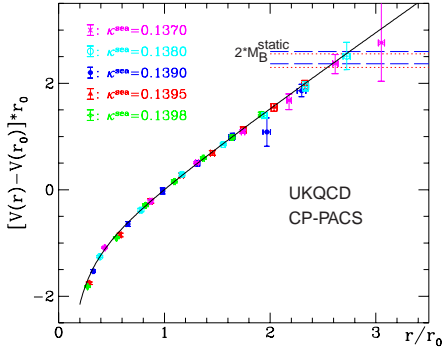}
\end{center}
\vskip -0.7in
\caption{Dynamical fermion loop effects on the static potential.}
\label{fig:fig05f}
\end{figure}

\begin{figure}[htb]
\begin{center}
\epsfysize=2.5in\epsfbox[74 242 477 695]{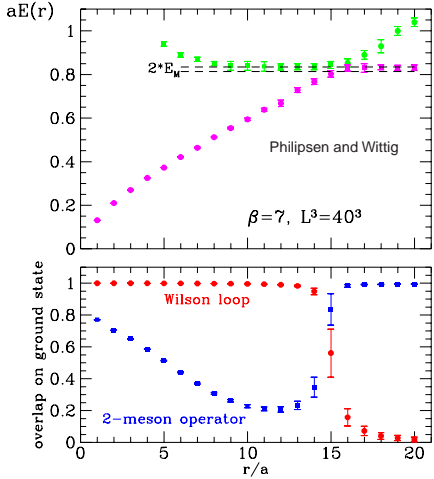}
\end{center}
\vskip -0.7in
\caption{Mixing effects between the Wilson loop operator and two
static mesons.}
\label{fig:fig06f}
\end{figure}

In QCD simulations with two flavors of dynamical quarks, where the
crossover range has been reached, 
new results were reported at the conference (Fig.\ref{fig:fig05f})
without any visible string breaking effects~\cite{Kenway,UKQCD,PACS}.
It was suggested using strong coupling 
ideas~\cite{Drummond} that string
breaking is a mixing phenomenon, involving both the string and the
static two-meson state. Thus, in order to confirm the mixing picture, the
conventionally used Wilson loops have to be supplemented by explicit
two-meson operators. This idea was illustrated in
two contributions to
the conference~\cite{Wittig,Sommer1} 
studying the Georgi-Glashow model with the Higgs
field in the fundamental representation playing
the role of the screening matter field. In Fig.\ref{fig:fig06f}
results from the three-dimensional model are shown~\cite{Wittig}
with findings very similar to the four-dimensional 
study~\cite{Sommer1}.
Since the quenched static potential between adjoint sources is
screened by the gluon field, its study would also provide further tests
on string breaking~\cite{Campbell,Trot2}.

It was reported at the conference 
from the study of Polyakov loop correlators, where mixing
with the static meson pair is different from Wilson loops,
that the finite temperature static potential does exhibit
screening behavior~\cite{Detar}. Since the calculation was done
just below the deconfining temperature it would be interesting
to extend this work to lower temperatures.
It was also reported that the screening of large Wilson
loops was easy to detect in lower dimensional models~\cite{Duncan,Trot3}.

\section*{Acknowledgements}
I would like to thank G.~Bali, R.~Burkhalter, R.~Kenway,
P.~Lacock, C.~Morningstar, M.~Peardon, and D.~Toussaint for help
in the preparation of this talk. I am also thankful
to the organizers of the conference who created a stimulating
atmosphere throughout the meeting.
This work was supported by the U.S.~DOE, Grant No.\ DE-FG03-97ER40546.

\end{document}